\documentclass[pra,aps,twocolumn,10pt,showpacs,groupedaddress,superscriptaddress,floatfix,notitlepage]{revtex4-2}
\usepackage{amsmath,amsfonts,amssymb,graphics,graphicx,epsfig,color,times}
\usepackage[utf8x]{inputenc}
\usepackage{color}
\usepackage{bbm, dsfont}
\usepackage{subfigure}
\usepackage{hyperref}
\usepackage{mathrsfs}
\usepackage{verbatim}
\begin{document}
\title{State carving in a chirally-coupled atom-nanophotonic cavity}

\author{W. S. Hiew}
\affiliation{Department of Physics, National Taiwan University, Taipei 10617, Taiwan}

\author{H. H. Jen}
\email{sappyjen@gmail.com}
\affiliation{Institute of Atomic and Molecular Sciences, Academia Sinica, Taipei 10617, Taiwan}
\affiliation{Physics Division, National Center for Theoretical Sciences, Taipei 10617, Taiwan}

\date{\today}
\renewcommand{\r}{\mathbf{r}}
\newcommand{\f}{\mathbf{f}}
\renewcommand{\k}{\mathbf{k}}
\def\p{\mathbf{p}}
\def\q{\mathbf{q}}
\def\bea{\begin{eqnarray}}
\def\eea{\end{eqnarray}}
\def\ba{\begin{array}}
\def\ea{\end{array}}
\def\bdm{\begin{displaymath}}
\def\edm{\end{displaymath}}
\def\red{\color{red}}
\pacs{}
\begin{abstract}
Coherent quantum control of multiqubit systems represents one of the challenging tasks in quantum science and quantum technology. Here we theoretically investigate the reflectivity spectrum in an atom-nanophotonic cavity with collective nonreciprocal couplings. In the strong-coupling regime with a high cooperativity, we theoretically predict distinct on-resonance spectral dips owing to destructive interferences of chiral couplings. Due to the well-separated multiple dips in the spectrum, a contrasted reflectivity suggests a new control knob over the desired entangled state preparation in the basis of coupled and uncoupled states from the atoms' internal hyperfine ground states. We propose to utilize such atom-nanophotonic cavity to quantum engineer the atomic internal states via photon-mediated dipole-dipole interactions in the coupled state and the chirality of decay channels, where the atomic Bell state and W states for arbitrary number of atoms can be tailored and heralded by state carving in the single-photon reflection spectrum. Our results pave the way toward quantum engineering of multiqubit states and offer new opportunities for coherent and scalable multipartite entanglement transport in atoms coupled to nanophotonic devices.
\end{abstract}
\maketitle
\section{Introduction}

Quantum state engineering aims to prepare and manipulate entangled states with high controllability. The generated entanglement among various platforms, neutral atoms, solid-state qubits or optical photons, has been essential as a resource for many central applications in quantum science and quantum technology. Owing to the controlled interactions between atoms and photons \cite{Hammerer2010, Reiserer2015}, an unparalleled supremacy of quantum systems holds promise to a quantum revolution that classical counterparts cannot foresee. Recently, the cavity carving protocol \cite{Sorensen2003, Chen2015, Welte2017} in cavity-based quantum networks \cite{Reiserer2015, Welte2018, Thomas2022} has been extended to an atom-nanophotonic interface \cite{Chang2018, Samutpraphoot2020, Sheremet2021}, where an entangled atomic Bell state can be prepared and heralded upon projective single-photon measurements by carving or removing the unwanted state components. This atom-nanophotonic interface allows strong light-atom couplings owing to the confined and guided light modes \cite{Vetsch2010, Thompson2013, Goban2015, Corzo2019, Kim2019}, in which a transportable entanglement can be envisioned in a scalable setup \cite{Dordevic2021}.   

Furthermore, the couplings between quantum emitters in an atom-nanophotonic interface can be tailored to be directional or nonreciprocal \cite{Mitsch2014, Bliokh2014, Ramos2014, Pichler2015, Lodahl2017} via external magnetic fields, where the superradiant emissions from these emitters manifest an infinite-range nature of the resonant dipole-dipole interactions \cite{Solano2017}. This collective and long-range dipole-dipole interaction underlies the distinct radiative dynamics \cite{Albrecht2019, Jen2020_subradiance, Pennetta2022, Pennetta2022_2}, strongly correlated photons \cite{Mahmoodian2018} or atoms \cite{Jen2020_steady, Jen2020_disorder, Jen2022_correlation}, the atom-photon bound states \cite{Mahmoodian2020, Kim2021, Jen2021_bound}, and superior cooling behaviors in trapped atoms \cite{Chen2023, Wang2022, Wang2023}. A controlled and tunable chiral coupling has yet been employed or investigated in an atom-nanophotonic cavity, where collective light-atom couplings can lead to strong atom-atom correlations and unexplored parameter regimes can arise for new applications in quantum technology. 

In this article, we study the single-photon reflectivity spectrum in an atom-nanophotonic cavity with many atoms. In the strong-coupling regime, we obtain distinct spectral dips in the spectrum, which can be attributed to interferences of chiral couplings. In this platform, we propose to use state carving technique to quantum engineer an atomic Bell or atomic W states via photon-mediated dipole-dipole interactions, which provides the foundation to coherent and scalable entanglement transport. The remainder of the paper is organized as follows. In Sec. II, we introduce the model of a chirally-coupled atom-nanophotonic cavity, where the directionality of couplings can be tunable. In Sec. III, we provide the analytical formula for the single-photon reflectivity for the case of two atoms. In Sec. IV, we present the results of chiral-coupling-induced dip in the reflectivity spectrum for different single-atom cooperativities. We then describe the state carving protocol for atomic W states in Sec. V. Finally, we discuss and conclude in Sec. VI.   

\section{A chirally-coupled atom-nanophotonic cavity}

\begin{figure}[t]
\centering
\includegraphics[width=0.48\textwidth]{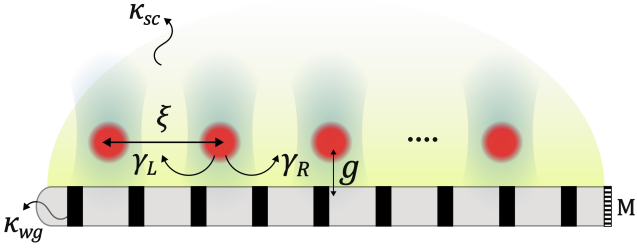}
\caption{A schematic plot of an atom-nanophotonic cavity. An equidistant atoms in an optical tweezer array with an interparticle distance $\xi\lambda/(2\pi)$ couple to the cavity with an atom-photon coupling constant $g$. Mediated by the evanescent waves and the propagating modes supported on the nanophotonic waveguide, effective nonreciprocal decay channels can be described by the left-propagating $\gamma_{\rm L}$ and right-propagating decay rate $\gamma_{\rm R}$ with a total decay rate $\gamma=\gamma_{\rm L}+\gamma_{\rm R}$. $\kappa_{\rm wg}$ and $\kappa_{\rm sc}$ represent the cavity decay rate to the waveguide and the nonguided rate for scattered light, respectively. `M' stands for a mirror with total reflection in this single-sided cavity, from which a single-photon reflective measurement can be conducted through the waveguide.}\label{fig1}
\end{figure}

We consider an atom-nanophotonic cavity system as shown schematically in Fig. \ref{fig1}. The atoms in an optical tweezer array \cite{Barredo2016, Endres2016, Kim2016, Barredo2018, Mello2019, Brooks2021, Sheng2021, Sheng2022, Huft2022} can be placed close to the nanophotonic cavity to allow a strong-coupling regime \cite{Samutpraphoot2020}. Under a total atomic spontaneous decay rate $\gamma$ composed of nonreciprocal decay channels $\gamma_{\rm L(R)}$, atom-photon coupling $g$, and a total cavity decay rate $\kappa=\kappa_{\rm wg}+\kappa_{\rm sc}$ involving the cavity decay rate to the waveguide (wg) and the nonguided rate for scattered (sc) light, respectively, a single-atom cooperativity can be defined as $C\equiv 4g^2/(\kappa\gamma)$ which characterizes a strong-coupling regime when $C\gg 1$.  

In addition to high cooperativity, the strongly-coupled atom-waveguide system with dominant chiral couplings has been experimentally realized in superconducting qubits \cite{Roushan2017, Wang2019}, quantum dots \cite{Luxmoore2013, Arcari2014, Yalla2014, Sollner2015}, and trapped atoms \cite{Tiecke2014, Mitsch2014, Solano2017, Corzo2019}, where the chiral couplings with a tunable directionality can be controlled under external magnetic fields \cite{Mitsch2014, Lodahl2017}. A useful measure of $\beta=(\gamma_{\rm L}+\gamma_{\rm R})/(\gamma_{\rm L}+\gamma_{\rm R}+\gamma_{\rm nr})$ \cite{Lodahl2017} can quantify how close the system is to the strong coupling regime, where $\gamma_{\rm nr}$ represents the nonradiative loss other than the propagating modes. $\beta$ has been reported more than $90\%$ in the atom-nanofiber system \cite{Mitsch2014} and more than $98\%$ in the quantum dot in the photonic crystal waveguide \cite{Arcari2014, Sollner2015}. This coupling can be further enhanced by placing quantum emitters on the surface of an optical nanofiber \cite{Yalla2014} or by using an external cavity \cite{Tiecke2014}. Therefore, before we introduce the theoretical model in the below, we note that the dominant atomic decay channels can be fully determined by $\gamma_{\rm L}$ and $\gamma_{\rm R}$, which should be valid under the strong coupling regime, that is $\beta\approx 1$. Throughout the paper, we focus on the regime with a high $C$ and near-unity $\beta$ in the atom-nanophotonic cavity. 

The dynamics of this atom-photon coupling system (atomic dipole operators $\sigma_\mu^\dag\equiv|e\rangle_\mu\langle g|$ for the ground state $|g\rangle$ and the excited state $|e\rangle$, and cavity photon operator $a$) can be described by the Heisenberg-Langevin equation, which for an arbitrary operator $A$ reads,
\bea
\dot{A}(t)=&&i[H_{s}+H_{\rm L}+H_{\rm R},A]+\gamma_{\rm L}D[c_{\rm L}]A+\gamma_{\rm R}D[c_{\rm R}]A\nonumber\\
&&+\kappa D[a]A. \label{A}
\eea
The coherent components for the atom-photon coupling Hamiltonian $H_s$ and the collective energies $H_{\rm L(R)}$ from photon-mediated dipole-dipole interactions are ($\hbar$ $=$ $1$)  
\bea
H_{s}=&&\omega_{c}a^{\dagger}a+\omega_a\sum_{\mu=1}^N\sigma_\mu^{\dagger}\sigma_\mu\nonumber\\
&&+\sum_{\mu=1}^N g\left(e^{-ik_sx_\mu}a^{\dagger}\sigma_\mu+e^{ik_sx_\mu}\sigma_\mu^{\dagger}a\right), \\
H_{\rm L(R)} =&& -i\frac{\gamma_{\rm L(R)}}{2} \sum_{\mu<(>)\nu}^N\left(e^{ik_s|x_\mu-x_\nu|} \sigma_\mu^\dag\sigma_\nu-\textrm{H.c.}\right),
\eea
where $\omega_{c}$ and $\omega_a$ are the resonance frequencies of the cavity and the atoms, respectively. The atomic positions can be ordered as $x_1$ $<$ $x_2$ $<...<$ $x_{N-1}$ $<$ $x_N$ for $N$ coupled atoms, and $k_s$ denotes the wave vector of the guided mode. The dissipative parts in Eq. (\ref{A}) are defined as $D[c]A\equiv c^\dag Ac-\{c^\dag c,A\}/2$ with $c_{\rm L}\equiv\sum_{\mu=1}^N e^{ik_sx_\mu}\sigma_\mu$ and $c_{\rm R}=\sum_{\mu=1}^N e^{-ik_sx_\mu}\sigma_\mu$ \cite{Pichler2015} for atoms and with $c=a$ for cavity photons. The associated $\gamma_{\rm L(R)}$ quantify the nonreciprocal decay rates to the left (L) or the right (R) of the atomic array, which are normalized by the total decay rate $\gamma\equiv 2|dq(\omega)/d\omega|_{\omega=\omega_a}g_{k_s}^2L$ \cite{Tudela2013}, with the inverse of group velocity $|dq(\omega)/d\omega|$, a resonant wave vector $q(\omega)$, the coupling strength $g_{k_s}$, and the quantization length $L$. For an equidistant array of atoms, a dimensionless $\xi\equiv k_s |x_{\mu+1}-x_{\mu}|$ carries the information of an interparticle distance and intends to characterize the strength of the collective dipole-dipole interactions. 

We note that the effective collective dipole-dipole interactions described by $H_{\rm L(R)}$ can be obtained by introducing the one-dimensional reservoir modes of the waveguide \cite{Pichler2015}, which interact with the atoms via the evanescent waves, and then tracing them out. This leads to the origin of collective spin-exchange interactions between the constituent atoms and a resource for exotic many-body dynamics and improved photon storage fidelities \cite{Garcia2017}. 

\section{Analytical form of the single-photon reflectivity}

Before we introduce the state carving protocol, we present the case of two atoms coupled with a nanophotonic cavity under nonreciprocal couplings, as the foundation for multiqubit state carving. We obtain the equations of motion of $A\in\{a,\sigma_1,\sigma_2\}$ based on the input-output formalism \cite{Caneva12015} in a weak excitation limit ($|g\rangle_\mu\langle g|\approx 1$) with $a_{\rm out}+a_{\rm in}=\sqrt{\kappa_{\rm wg}}a$, 
\bea
\dot{a}(t)=&&i\delta_{c}a-ig\sum_{\mu=1}^N e^{-ik_s x_\mu}\sigma_\mu-\frac{\kappa}{2}a+\sqrt{\kappa_{\rm wg}}a_{\rm in}, \\
\dot{\sigma}_{1(2)}(t)=&&\left(i\delta-\frac{\gamma}{2}\right)\sigma_{1(2)}-\gamma_{\rm L(R)}e^{i\xi}\sigma_{2(1)}-ig e^{ik_s x_{1(2)}}a,\nonumber\\
\eea
where $a_{\rm in}$ is the input field, and $\delta_c\equiv \omega-\omega_c$ and $\delta\equiv \omega-\omega_a$ characterize the cavity and probe detunings, respectively. We note that the quantum noise terms have been excluded since only the observables with normal orders are concerned here. Without loss of generality, we set $\delta_c=\delta$, and the steady-state solutions of $A$ can be obtained self-consistently without higher-order terms of atom-atom correlations from more than a single excitation. The reflectivity can be calculated as $r=a_{\rm out}/a_{\rm in}=\sqrt{\kappa_{\rm wg}}a/a_{\rm in}-1$ from the steady-state solutions, which becomes 
\bea
\frac{\kappa_{\rm wg}}{r+1}=\left\{\left(\frac{\kappa}{2}-i\delta\right)-\frac{g^{2}[2(i\delta-\frac{\gamma}{2})+(e^{i2\xi}\gamma_{L}+\gamma_{R})]}{\left(i\delta-\frac{\gamma}{2}\right)^{2}-e^{i2\xi}\gamma_{L}\gamma_{R}}\right\}.\nonumber\\\label{R}
\eea

\begin{figure}[t]
\centering
\includegraphics[width=0.48\textwidth]{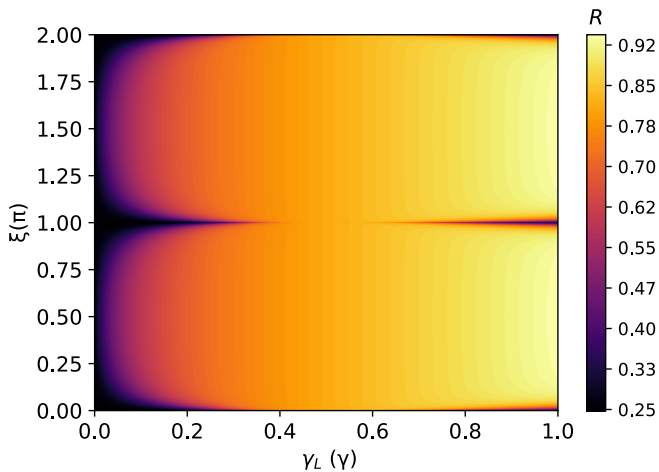}
\caption{On-resonance reflectivity spectrum $R=|r|^2$ at $\delta=0$ for various $\xi$ and $\gamma_{\rm L}$. At $\xi=n\pi$ with an integer $n$ or $\gamma_{\rm L}=0$, $\gamma/2$, $R$ remains constant with respect to all $\gamma_{\rm L}$ or $\xi$, respectively, and the lowest in the considered parameter regions other than $\gamma_{\rm L}=\gamma/2$. The maximal $R$ emerges at $\gamma_{\rm L}=\gamma$ when $\xi=\pi/2$ and $3\pi/2$. The parameters are chosen as $(\kappa_{\rm wg}, \kappa_{\rm sc}, g)/\gamma=(100,300,20)$ with $\gamma=2\pi\times 6$ MHz for rubidium atoms as an example, giving $C=4$. These adaptable reflection properties demonstrate a functionality that can be employed for system characterizations and state manipulations.}\label{fig2}
\end{figure}

The reflection spectrum $R$ can then be calculated as $|r|^2$ and can be directly obtained in the single-photon measurements in a single-sided cavity as shown in Fig. \ref{fig1}. Equation (\ref{R}) presents an intriguing interference effect of $e^{i2\xi}$, which associates with chiral couplings and reflects the collective and nonreciprocal spin-exchange effect. This result reduces to the case of two independent atoms when $\gamma_{\rm L(R)}\rightarrow 0$ \cite{Samutpraphoot2020}. In Fig. \ref{fig2}, we plot the on-resonance reflectivity spectrum $R$ from Eq. (\ref{R}), where versatile reflection properties can be controlled by tailoring the strengths of $\gamma_{\rm L}$ or $\xi$. We attribute this variation to the constructive or destructive interferences between the nonreciprocal couplings and photon-mediated dipole-dipole interactions, where nontrivial effect from collective and chiral spin-exchange processes arises. 

In Fig. \ref{fig2}, the reflection spectrum $R$ shows a period of mutual atomic distance $\xi=\pi$ owing to the factor of $e^{i2\xi}$. As a comparison to the on-resonance case of two independent atoms of $r_{\rm ind}=2\kappa_{\rm wg}[1+2C]^{-1}/\kappa-1$ \cite{Samutpraphoot2020} where $|r_{\rm ind}|^2=(17/18)^2\approx 0.89$, we have multiple parameter regimes which vary significantly from the independent case. First of all, the minimums in Fig. \ref{fig2} emerge as $r=2\kappa_{\rm wg}/\kappa-1$ at $\xi=\pi$ for all $\gamma_{\rm L}$ and at $\gamma_{\rm L}=0$ for all $\xi$, where $R=0.25$. In huge contrast to the independent case, these parameter regimes behave as if there are no atoms coupling to the cavity, showing a purely cavity effect, and present a destructive interference that leads to a disappearance of the cooperativity $C$ from the atom-photon couplings. On the other hand, along the line at $\xi=\pi/2$ in Fig. \ref{fig2}, we obtain $r_{\rm d}=2\kappa_{\rm wg}[1+4C]^{-1}/\kappa-1$ with $|r_{\rm d}|^2\approx 0.94$ for a directional coupling $\gamma_{\rm L}=\gamma$ and $r_{\rm rec}=2\kappa_{\rm wg}[1+C]^{-1}/\kappa-1$ with $|r_{\rm rec}|^2=0.81$ for a reciprocal coupling $\gamma_{\rm L}=\gamma_{\rm R}=\gamma/2$, respectively. The former case indicates an enhancement effect of single-atom cooperativity $C$, leading to a higher value of $R$, while the latter shows a suppressed effect of $C$. This manifests a constructive and a destructive interference on the reflection spectrum, respectively, by tuning $\gamma_{\rm L}$. These regimes can be distinguished when a moderate $C$ is considered, and all three scenarios of $r_{\rm ind}$, $r_{\rm d}$, and $r_{\rm rec}$ lead to a near-perfect reflection when $C\gg 1$. In other words, the single-photon reflectivity in an atom-nanophotonic cavity with a mild cooperativity $C$ can be used to characterize the mutual atomic distance $\xi\lambda/(2\pi)$ associated with the collective photon-mediated dipole-dipole interactions and to reveal the associated atom-atom correlations that are crucial for laser cooling of atoms \cite{Xu2016, Chen2023}.

\begin{figure}[b]
\centering
\includegraphics[width=0.48\textwidth]{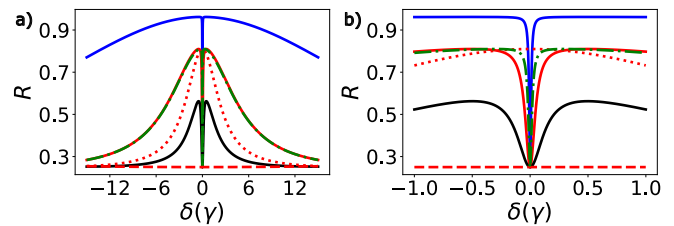}
\caption{Chiral-coupling-induced dip in the reflectivity spectrum at $\gamma_L=\gamma$ with $\xi=n\pi$. (a) Single-photon reflectivity spectrum $R$ is shown at $(\kappa, g)/\gamma=(400,50)$ (solid-blue line), $(400,20)$ (solid-red line), $(400,10)$ (solid-black line) for $N=2$, which corresponds to $C=25$, $4$, and $1$, respectively. Several comparisons can be made to the single-atom case for $(\kappa, g)/\gamma=(400,20)$ (dotted-red line), the case without atoms as a baseline for $(\kappa, g)/\gamma=(400,0)$ (dashed-red line), and the case of $\gamma_L/\gamma=0.8$ for $(\kappa, g)/\gamma=(400,20)$ (dash-dotted green line). The $\kappa_{\rm sc}$ is chosen as three times of $\kappa_{\rm wg}$ as in Fig. \ref{fig2}. (b) A magnified plot for the reflectivity dips around the resonance at $\delta\approx 0$.}\label{fig3}
\end{figure}

\section{Chiral-coupling-induced reflectivity dip}

Next we focus on the directional coupling regime when $\gamma_{\rm L}=\gamma$ at $\xi=n\pi$ in Fig. \ref{fig2}, which we will employ to demonstrate the state carving protocol. In this regime, the most significant reduction in $R$ shows up as $\xi$ varies around an integer of $\pi$. By contrast, this reduction becomes less apparent and featureless as $\gamma_{\rm L}$ approaches the reciprocal coupling regime. A decrease in $R$ in an atom-nanophotonic cavity suggests an effective destructive interference from directional spontaneous emissions, which is evident in Fig. \ref{fig3} where we find spectral dips in the single-photon reflectivity. These spectral structures sustain and are symmetric with respect to $\gamma_{\rm L}=\gamma/2$ at $\xi=n\pi$, which can be seen as well in Fig. \ref{fig2}. Remarkably, at $\gamma_{\rm L}=0$ ($\gamma_{\rm R}=\gamma$), the single-photon reflection preserves for arbitrary $\xi$ and behaves exactly the same as in the directional coupling regime of $\gamma_{\rm L}=\gamma$ with $\xi=n\pi$. This provides the most robust operating regime against the atomic position fluctuations, and this regime shares the essential results presented in Fig. \ref{fig3}.

First in Fig. \ref{fig3}(a) aside from the spectral dips, the effect of cooperativity reflects in the maximal value of $R$, which approaches $R\approx 1$ as $C$ increases. The essence of cooperativity enhances the contrast of near-resonance reflectivity between the cases with and without atoms, which lays the foundation for carving unwanted states. Comparing the single-atom case, the reflectivity of two atoms coupled collectively with each other in the cavity manifests a spectral broadening, approximately twice of the single-atom case under the same $C$. Interestingly, the maximal value of near-resonance $R$ under the same $C$ is the same as the maximal and on-resonance $R$ in the single-atom case, which is in contrast to the case of independent atoms for $N>1$. This indicates that the effect of photon-mediated dipole-dipole interactions at $\xi=n\pi$ manifests exactly a spectral broadening in the reflectivity profile. Furthermore, a different nonreciprocal coupling strength of $\gamma_{\rm L}\neq \gamma/2$ does not evidently modify the overall profile but significantly narrow the spectral dip until it disappears at $\gamma_{\rm L}=\gamma/2$.  

\begin{figure}[b]
\centering
\includegraphics[width=0.48\textwidth]{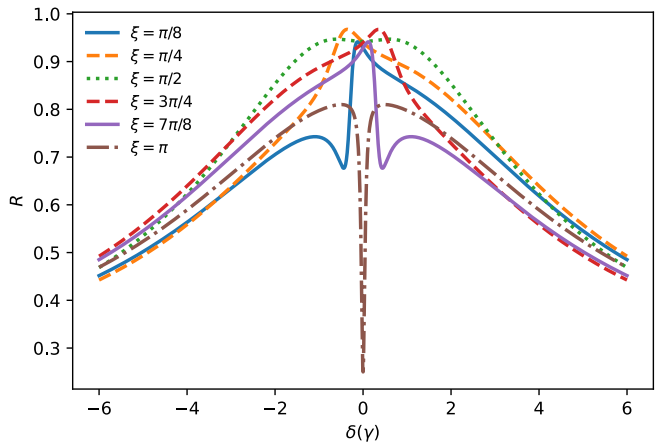}
\caption{Asymmetric spectral profiles for single-photon reflectivity. Various single-photon reflectivities are plotted for $\gamma_{\rm L}=\gamma$ at $\xi$ other than $n\pi$ or $(n+1/2)\pi$. An enhanced and maximal $R$ comparing the single-atom case emerges at deviated resonance conditions $\delta\neq 0$ with asymmetric profiles. The rest of system parameters are the same as in Fig. \ref{fig2}.}\label{fig3_2}
\end{figure}

In Fig. \ref{fig3}(b), we further zoom in the spectral dips for various cases of $C$ and $\gamma_{\rm L}$. We find a spectral narrowing when $C$ increases or when $\gamma_{\rm L}$ approaches $\gamma/2$, showing an interference between the collective spin-exchange coupling and its nonreciprocity. This resembles the electromagnetically-induced-transparency in an optical media \cite{Fleischhauer2005} or radiation-pressure-induced transparency in an optomechanical system \cite{Weis2010}, where a destructive quantum interference emerges between two excitation fields, leading to a transparent window for the probe field. By contrast, only a single-photon probe is present here in a single-sided cavity. The destructive interference renders the on-resonance reflectivity to the case without atoms, purely from an interplay between the cooperativity of the system, the chirality of the atomic decay rates, and the atomic relative distances. In Fig. \ref{fig3_2}, we present an asymmetric profile with or without a spectral dip and with an enhanced reflectivity at a deviated resonance condition when $\xi$ is chosen other than $n\pi$ or $(n+1/2)\pi$. The spectral profiles at $\xi<\pi/2$ and $\xi>\pi/2$ present a mirror symmetry with respect to $\xi=\pi/2$, which indicates a symmetry of $R$ as $\xi\rightarrow\xi-\pi$ and $\gamma_{\rm L}\rightarrow \gamma_{\rm R}$. These maximal $R$'s can even surpass the maximal values allowed in two independent atoms, showing an enhancement of $R$ from the constructive interference of chiral couplings.  

\begin{figure}[t]
\centering
\includegraphics[width=0.48\textwidth]{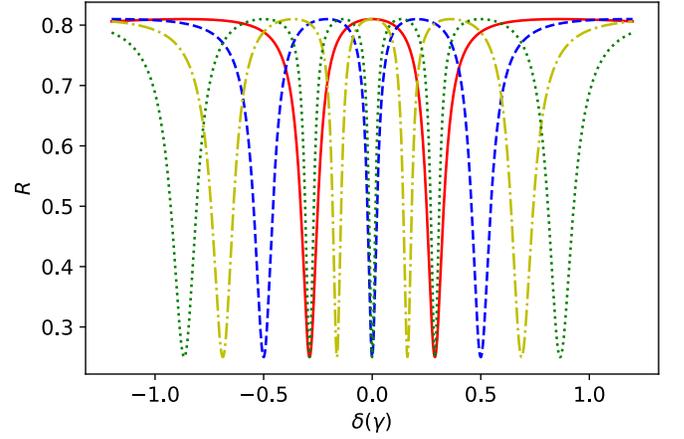}
\caption{The reflectivity spectrum for multiple spectral dips. The near-resonance spectral dips are plotted for $N=3,4,5,6$ (solid-red, dashed-blue, dash-dotted-yellow, dotted-green lines) at $\gamma_{\rm L}=\gamma$ and $\xi=n\pi$, with the same parameters of $(\kappa_{\rm wg}, \kappa_{\rm sc}, g)/\gamma=(100,300,20)$ and $C=4$ as in Fig. \ref{fig2}.}\label{fig4}
\end{figure}

This leads us to further investigate the interference effect from more atoms by numerically solving Eq. (\ref{A}), where $A\in\{a,\sigma_{1},\sigma_{2},\cdots,\sigma_{N}\}$ is sufficient under the weak field excitation limit. In Fig. \ref{fig4}, we demonstrate the cases for multiple equidistant atoms coupled to a nanophotonic cavity as an extension to Fig. \ref{fig3}(b). As $N$ increases, the overall spectral widths are broadened with $N-1$ narrow spectral dips in the reflectivity spectrum. This leads to an observation that the on-resonance spectral peaks (dips) emerge for odd (even) number of atoms, which exhibits the multiatom interferences and gives rise to a new pathway of controlling the atomic states via single-photon measurements.  

\section{State carving protocol for atomic W states}

Finally we propose a state carving protocol exploiting the reflectivity spectrum in a chirally-coupled atom-nanophotonic cavity. In the same spirit of cavity carving protocol \cite{Welte2017}, we consider $5S_{1/2}\rightarrow 5P_{3/2}$ transition in rubidium atoms and label the hyperfine ground states of $F=2$ and $F=1$ manifolds as coupled and uncoupled to the cavity \cite{Samutpraphoot2020}, which we denote them as $|1\rangle$ and $|0\rangle$, respectively. For a total number of atoms $M=2$, the atoms can both be prepared initially in $|00\rangle$. After applying a global $\pi/2$ rotation on the atoms, the atoms evolve to $\frac{1}{2}(\left|11\right\rangle -\left|01\right\rangle -\left|10\right\rangle +\left|00\right\rangle )$. When we send a weak and resonant pulse into the atom-cavity system, the on-resonance reflectivity $R$ in Fig. \ref{fig3} shows a low $R$ for $N=0,2$ but a high $R$ for $N=1$, which effectively carves or removes the states $|00\rangle$ ($N=0$ as two uncoupled atoms) and $|11\rangle$ ($N=2$ as two coupled atoms). This singles out the atomic Bell state with a global phase, $(|01\rangle +|10\rangle)/\sqrt{2}$, under a high-probability and heralded preparation upon projective single-photon measurements.  

We further extend the protocol to prepare the atomic W states with $M\geq 3$ by utilizing the chiral-coupling-induced spectral dips predicted in Fig. \ref{fig4}. After the similar cavity carving protocol for $M=2$, we are left with the single and triple coupled states for $M=3$ as an example. To remove the unwanted state $|111\rangle$, we can apply a second projective single-photon measurement with a photon detuning that operates at the minimal $R$ for $N=3$, which can be done in the isolated spectral dips as shown in Fig. \ref{fig4} at $\delta\approx \pm 0.3\gamma$. Under the second weak pulse measurement at this photon detuning, the whole two-stage protocol creates an atomic W state with only a single coupled state manifold, $(|100\rangle+|010\rangle+|001\rangle)/\sqrt{3}$, where the double coupled states with a high $R$ at this detuning are already carved in the first stage. Essentially, a general atomic W state, $\sum_{m=1}^M(|1\rangle_m\langle 0|)|0\rangle^{\otimes M}/\sqrt{M}$, can be prepared by multiple projective measurements ($M/2$ or $(M+1)/2$ of them for even or odd $M$ quantum registers with $M\geq 2$) by operating the weak pulses at various photon detunings with a minimal $R$ sequentially, in a sense to carve the unwanted components with higher number occupations in the coupled states.     

\section{Discussion and conclusion}

Our new protocol for state carving in a chirally-coupled atom-nanophotonic cavity remains at least two challenges. One is the requirement of strong-coupling regime with high enough cooperativity, which requires the trapped atoms in an optical tweezer array close enough to the waveguide. Many ongoing efforts are under active developments to reach an efficient coupling in the atom-waveguide interfaces \cite{Arcari2014, Tiecke2014, Yalla2014, Sollner2015}. The other relates to the heralding efficiency of atomic W states under a series of projective measurements. This can be overcome by adopting a high-cooperativity coupled system, which allows more contrasted on- or near-resonance reflectivities comparing to the baseline without atoms. 

The dips obtained in this work are purely collectively interfered from the chiral couplings and photon-mediated dipole-dipole interactions, in huge contrast to the vacuum induced transparency \cite{Suzuki2011} where its transparency window is opened by a strong coupling to the cavity photon frequency detuned from the probe photon. Furthermore, our state-carving model utilizes a cavity resonance in accordance with the probe photon which probes the atom in the coupled state. In the vacuum induced transparency, a single atom can host this transparency window opening, whereas in our consideration in the atom-nanophotonic cavity, the reflection dips emerge only when multiple atoms of $N\geq 2$ are involved, that is when the collective spin-exchange couplings between the atoms are present. 

In conclusion, we theoretically obtain the intriguing reflectivity spectrum in an atom-nanophotonic cavity with collective and nonreciprocal photon-mediated couplings. In the strong-coupling regime, we present the on-resonance and near-resonance spectral dips in the single-photon reflectivity owing to the destructive interferences of chiral couplings. We propose to generate an atomic Bell state and atomic W states by projective and heralded single-photon measurements, which can be realized in our proposed state carving protocol. Our results open new opportunities to generate Greenberger-Horne-Zeilinger states with an assistance of controlled-NOT gate and to synthesize multipartite cluster states as in atom-cavity-based networks \cite{Thomas2022}. 

\section*{ACKNOWLEDGMENTS}
We acknowledge support from the Ministry of Science and Technology (MOST), Taiwan, under the Grant No. MOST-109-2112-M-001-035-MY3 and No. MOST-111-2119-M-001-002. We are also grateful for support from TG 1.2 and TG 3.2 of NCTS at NTU. 


\begin{thebibliography}{99}
\bibitem{Hammerer2010} K. Hammerer, A. S. S\o{}rensen, and E. S. Polzik, Quantum interface between light and atomic ensembles, Rev Mod Phys. {\bf 82}, 1041 (2010). 
\bibitem{Reiserer2015} A. Reiserer and G. Rempe, Cavity-based quantum networks with single atoms and optical photons, Rev. Mod. Phys. {\bf 87}, 1379 (2015).
\bibitem{Sorensen2003} A. S. Sørensen and K. Mølmer, Probabilistic Generation of Entanglement in Optical Cavities, Phys. Rev. Lett. {\bf 90}, 127903 (2003). 
\bibitem{Chen2015} W. Chen, J. Hu, Y. Duan, B. Braverman, H. Zhang, and V. Vuletić, Carving Complex Many-Atom Entangled States by Single-Photon Detection, Phys. Rev. Lett. {\bf 115}, 250502 (2015).
\bibitem{Welte2017} S. Welte, B. Hacker, S. Daiss, S. Ritter, and G. Rempe, Cavity Carving of Atomic Bell States, Phys. Rev. Lett. {\bf 118}, 210503 (2017).
\bibitem{Welte2018} S. Welte, B. Hacker, S. Daiss, S. Ritter, and G. Rempe, Photon-Mediated Quantum Gate between Two Neutral Atoms in an Optical Cavity, Phys. Rev. X {\bf 8}, 011018 (2018).
\bibitem{Thomas2022} P. Thomas, L. Ruscio, O. Morin, and G. Rempe, Efficient generation of entangled multiphoton graph states from a single atom, Nature {\bf 608}, 677 (2022). 
\bibitem{Chang2018} D. E. Chang, J. S. Douglas, A. Gonz\'alez-Tudela, C.-L. Hung, H. J. Kimble, Colloquium: Quantum matter built from nanoscopic lattices of atoms and photons, Rev. Mod. Phys. {\bf 90}, 031002 (2018).
\bibitem{Samutpraphoot2020} P. Samutpraphoot, T. Dordevi\ifmmode \acute{c}\else \'{c}\fi{}, P. L. Ocola, H. Bernien, C. Senko, V. Vuletić, and M. D. Lukin, Strong Coupling of Two Individually Controlled Atoms via a Nanophotonic Cavity, Phys. Rev. Lett. {\bf 124}, 063602 (2020). 
\bibitem{Sheremet2021} A. S. Sheremet, M. I. Petrov, I. V. Iorsh, A. V. Poshakinskiy, A. N. Poddubny, Waveguide quantum electrodynamics: collective radiance and photon-photon correlations, Rev. Mod. Phys. {\bf 95}, 015002 (2023). 
\bibitem{Vetsch2010} E. Vetsch, D. Reitz, G. Sagu\'e, R. Schmidt, S. T. Dawkins, and A. Rauschenbeutel, Optical Interface Created by Laser-Cooled Atoms Trapped in the Evanescent Field Surrounding an Optical Nanofiber, Phys. Rev. Lett. {\bf 104}, 203603 (2010).
\bibitem{Thompson2013} J. D. Thompson, T. G. Tiecke, N. P. de Leon, J. Feist, A. V. Akimov, M. Gullans, A. S. Zibrov, V. Vuletić, and M. D. Lukin, Coupling a Single Trapped Atom to a Nanoscale Optical Cavity, Science {\bf 340}, 1202 (2013).
\bibitem{Goban2015} A. Goban, C.-L. Hung, J. D. Hood, S.-P. Yu, J. A. Muniz, O. Painter, and H. J. Kimble, Superradiance for Atoms Trapped along a Photonic Crystal Waveguide, Phys.Rev.Lett. {\bf 115}, 063601 (2015).
\bibitem{Corzo2019} N. V. Corzo, J. Raskop, A. Chandra, A. S. Sheremet, B. Gouraud, and J. Laurat, Waveguide-coupled single collective excitation of atomic arrays, Nature {\bf 566}, 359 (2019).
\bibitem{Kim2019} M. E. Kim, T.-H. Chang, B. M. Fields, C.-A. Chen, and C.-L. Hung, Trapping single atoms on a nanophotonic circuit with configurable tweezer lattices, Nat. Commun. {\bf 10}, 1647 (2019).
\bibitem{Dordevic2021} T. Dordevi\ifmmode \acute{c}\else \'{c}\fi{}, P. Samutpraphoot, P. L. Ocola, H. Bernien, B. Grinkemeyer, I. Dimitrova, V. Vuletić, and M. D. Lukin, Entanglement transport and a nanophotonic interface for atoms in optical tweezers, Science {\bf 373}, 1511 (2021). 
\bibitem{Mitsch2014} R. Mitsch, C. Sayrin, B. Albrecht, P. Schneeweiss, and A. Rauschenbeutel, Quantum state-controlled directional spontaneous emission of photons into a nanophotonic waveguide, Nat. Commun. {\bf 5}, 5713 (2014).
\bibitem{Bliokh2014} K. Y. Bliokh, A. Y. Bekshaev, and F. Nori, Extraordinary momentum and spin in evanescent waves, Nat. Commun. {\bf 5}, 3300 (2014).
\bibitem{Ramos2014} T. Ramos, H. Pichler, A. J. Daley, and P. Zoller, Quantum Spin Dimers from Chiral Dissipation in Cold-Atom Chains, Phys. Rev. Lett. {\bf 113}, 237203 (2014).
\bibitem{Pichler2015} H. Pichler, T. Ramos, A. J. Daley, and P. Zoller, Quantum optics of chiral spin networks, Phys. Rev. A {\bf 91}, 042116 (2015).
\bibitem{Lodahl2017} P. Lodahl, S. Mahmoodian, S. Stobbe, A. Rauschenbeutel, P. Schneeweiss, J. Volz, H. Pichler, and P. Zoller, Chiral quantum optics, Nature {\bf 541}, 473 (2017).
\bibitem{Solano2017} P. Solano, P. Barberis-Blostein, F. K. Fatemi, L. A. Orozco, and S. L. Rolston, Super-radiance reveals infinite-range dipole interactions through a nanofiber, Nat. commun. {\bf 8}, 1857 (2017).
\bibitem{Albrecht2019} A. Albrecht, L. Henriet, A. Asenjo-Garcia, P. B Dieterle, O. Painter, and D. E. Chang, Subradiant states of quantum bits coupled to a one-dimensional waveguide, New J. Phys. {\bf 21}, 025003 (2019).
\bibitem{Jen2020_subradiance} H. H. Jen, M.-S. Chang, G.-D. Lin, and Y.-C. Chen, Subradiance dynamics in a singly excited chirally coupled atomic chain, Phys. Rev. A {\bf 101}, 023830 (2020).
\bibitem{Pennetta2022} R. Pennetta, M. Blaha, A. Johnson, D. Lechner, P. Schneeweiss, J. Volz, and A. Rauschenbeutel, Collective Radiative Dynamics of an Ensemble of Cold Atoms Coupled to an Optical Waveguide, Phys. Rev. Lett. {\bf 128}, 073601 (2022).
\bibitem{Pennetta2022_2} R. Pennetta, D. Lechner, M. Blaha , A. Rauschenbeutel, P. Schneeweiss, and J. Volz, Observation of Coherent Coupling between Super- and Subradiant States of an Ensemble of Cold Atoms Collectively Coupled to a Single Propagating Optical Mode, Phys. Rev. Lett. {\bf 128}, 203601 (2022).
\bibitem{Mahmoodian2018} S. Mahmoodian, M. \ifmmode \check{C}\else \v{C}\fi{}epulkovskis, S. Das, P. Lodahl, K. Hammerer, and A. S. S\o{}rensen, Strongly Correlated Photon Transport in Waveguide Quantum Electrodynamics with Weakly Coupled Emitters, Phys. Rev. Lett. {\bf 121}, 143601 (2018).
\bibitem{Jen2020_steady} H. H. Jen, Steady-state phase diagram of a weakly driven chiral-coupled atomic chain, Phys. Rev. Research {\bf 2}, 013097 (2020). 
\bibitem{Jen2020_disorder} H. H. Jen, Disorder-assisted excitation localization in chirally coupled quantum emitters, Phys. Rev. A {\bf 102}, 043525 (2020). 
\bibitem{Jen2022_correlation} H. H. Jen, Quantum correlations of localized atomic excitations in a disordered atomic chain, Phys. Rev. A {\bf 105}, 023717 (2022). 
\bibitem{Mahmoodian2020} S. Mahmoodian, G. Calajó, D. E. Chang, K. Hammerer, and A. S. Sørensen, Dynamics of Many-Body Photon Bound States in Chiral Waveguide QED, Phys. Rev. X {\bf 10}, 031011 (2020).
\bibitem{Kim2021} E. Kim, X. Zhang, V. S. Ferreira, J. Banker, J. K. Iverson, A. Sipahigil, M. Bello, A. Gonz\'alez-Tudela, M. Mirhosseini, and O. Painter, Quantum Electrodynamics in a Topological Waveguide, Phys. Rev. X {\bf 11}, 011015 (2021).
\bibitem{Jen2021_bound} H. H. Jen, Bound and subradiant multiatom excitations in an atomic array with nonreciprocal couplings, Phys. Rev. A {\bf 103}, 063711 (2021). 
\bibitem{Chen2023} C.-C. Chen, Y.-C. Wang, C.-C. Wang, and H. H. Jen, Chiral-coupling-assisted refrigeration in trapped ions, to be published in J. Phys. B.
\bibitem{Wang2022} C.-C. Wang, Y.-C. Wang, C.-H. Wang, C.-C. Chen, and H. H. Jen, Superior dark-state cooling via nonreciprocal couplings in trapped atoms, New. J. Phys. {\bf 24}, 113020 (2022). 
\bibitem{Wang2023} C.-H. Wang, Y.-C. Wang, C.-C. Chen, C.-C. Wang, H. H. Jen, Enhanced dark-state sideband cooling in trapped atoms via photon-mediated dipole-dipole interactions, Phys. Rev. A {\bf 107}, 023117 (2023). 
\bibitem{Barredo2016} D. Barredo, S. de L\'{e}s\'{e}leuc, V. Lienhard, T. Lahaye, A. Browaeys, An atom-by-atom assembler of defect-free arbitrary two-dimensional atomic arrays, Science {\bf 354}, 1021 (2016).
\bibitem{Endres2016} M. Endres, H. Bernien, A. Keesling, H. Levine, E. R. Anschuetz, A. Krajenbrink, C. Senko, V. Vuletic, M. Greiner, M. D. Lukin, Atom-by-atom assembly of defect-free one-dimensional cold atom arrays, Science {\bf 354}, 1024 (2016).
\bibitem{Kim2016} H. Kim, W. Lee, H.-g. Lee, H. Jo, Y. Song, and J. Ahn, {\it In situ} single-atom array synthesis using dynamic holographic optical tweezers, Nat. Commun. {\bf 7}, 13317 (2016).
\bibitem{Barredo2018} D. Barredo, V. Lienhard, S. de L\'{e}s\'{e}leuc, T. Lahaye, and A. Browaeys, Synthetic three-dimensional atomic structures assembled atom by atom, Nature {\bf 561}, 79 (2018).
\bibitem{Mello2019} D. Ohl de Mello, D. Schäffner, J. Werkmann, T. Preuschoff, L. Kohfahl, M. Schlosser, and G. Birkl, Defect-Free Assembly of 2D Clusters of More Than $100$ Single-Atom Quantum Systems, Phys. Rev. Lett. {\bf 122}, 203601 (2019).
\bibitem{Brooks2021} R. V. Brooks, S. Spence, A. Guttridge, A. Alampounti, A. Rakonjac, L. A. McArd, J. M. Hutson, and S. L. Cornish, Preparation of one 87Rb and one 133Cs atom in a single optical tweezer, New J. Phys. {\bf 23}, 065002 (2021). 
\bibitem{Sheng2021} C. Sheng, J. Hou, X. He, P. Xu, K. Wang, J. Zhuang, X. Li, M. Liu, J. Wang, and M. Zhan, Efficient preparation of two-dimensional defect-free atom arrays with near-fewest sorting-atom moves, Phys. Rev. Research {\bf 3}, 023008 (2021). 
\bibitem{Sheng2022} C. Sheng, J. Hou, X. He, K. Wang, R. Guo, J. Zhuang, B. Mamat, P. Xu, M. Liu, J. Wang, and M. Zhan, Defect-Free Arbitrary-Geometry Assembly of Mixed-Species Atom Arrays, Phys. Rev. Lett. {\bf 128}, 083202 (2022). 
\bibitem{Huft2022} P. Huft, Y. Song, T. M. Graham, K. Jooya, S. Deshpande, C. Fang, M. Kats, and M. Saffman, Simple, passive design for large optical trap arrays for single atoms, Phys. Rev. A {\bf 105}, 063111 (2022). 
\bibitem{Roushan2017} P. Roushan, C. Neill, A. Megrant, Y. Chen, R. Babbush, R. Barends, B. Campbell, Z. Chen, B. Chiaro, A. Dunsworth, {\it et. al.}, Chiral ground-state currents of interacting photons in a synthetic magnetic field, Nat. Phys. {\bf 13}, 146 (2017).  
\bibitem{Wang2019} D.-W. Wang, C. Song, W. Feng, H. Cai, D. Xu, H. Deng, H. Li, D. Zheng, X. Zhu, H. Wang, {\it et. al.}, Synthesis of antisymmetric spin exchange interaction and chiral spin clusters in superconducting circuits, Nat. Phys. {\bf 15}, 382 (2019).  
\bibitem{Luxmoore2013} I. J. Luxmoore, N. A. Wasley, A. J. Ramsay, A. C. T. Thijssen, R. Oulton, M. Hugues, S. Kasture, V. G. Achanta, A. M. Fox, and M. S. Skolnick, Interfacing Spins in an InGaAs Quantum Dot to a Semiconductor Waveguide Circuit Using Emitted Photons, Phys. Rev. Lett. {\bf 110}, 037402 (2013). 
\bibitem{Arcari2014} M. Arcari, I. S\"{o}llner, A. Javadi, S. Lindskov Hansen, S. Mahmoodian, J. Liu, H. Thyrrestrup, E. H. Lee, J. D. Song, S. Stobbe, and P. Lodahl, Near-Unity Coupling Efficiency of a Quantum Emitter to a Photonic Crystal Waveguide, Phys. Rev. Lett. {\bf 113}, 093603 (2014).
\bibitem{Yalla2014} R. Yalla, M. Sadgrove, K. P. Nayak, and K. Hakuta, Cavity Quantum Electrodynamics on a Nanofiber Using a Composite Photonic Crystal Cavity, Phys. Rev. Lett. {\bf 113}, 143601 (2014).
\bibitem{Sollner2015} I. Söllner, S. Mahmoodian, S. L. Hansen, L. Midolo, A. Javadi, G. Kiršanskė, T. Pregnolato, H. El-Ella, E. H. Lee, J. D. Song, {\it et. al.}, Deterministic photon–emitter coupling in chiral photonic circuits, Nat. Nanotechnol. {\bf 10}, 775 (2015).  
\bibitem{Tiecke2014} T. G. Tiecke, J. D. Thompson, N. P. de Leon, L. R. Liu, V. Vuleti\'{c}, and M. D. Lukin, Nanophotonic quantum phase switch with a single atom, Nature {\bf 508}, 241 (2014).
\bibitem{Tudela2013} A. Gonz\'alez-Tudela and D. Porras, Mesoscopic Entanglement Induced by Spontaneous Emission in Solid-State Quantum Optics, Phys. Rev. Lett. {\bf 110}, 080502 (2013).
\bibitem{Garcia2017} A. Asenjo-Garcia, M. Moreno-Cardoner, A. Albrecht, H. J. Kimble, and D. E. Chang, Exponential Improvement in Photon Storage Fidelities Using Subradiance and “Selective Radiance” in Atomic Arrays, Phys. Rev. X {\bf 7}, 031024 (2017). 
\bibitem{Caneva12015} T. Caneva, M. T. Manzoni, T. Shi, J. S. Douglas, J. I. Cirac, and D. E. Chang, Quantum dynamics of propagating photons with strong interactions: a generalized input–output formalism, New J. Phys. {\bf 17}, 113001 (2015). 
\bibitem{Xu2016} M. Xu, S. B. J\"{a}ger, S. Sch\"{u}tz, J. Cooper, G. Morigi, and M. J. Holland, Supercooling of Atoms in an Optical Resonator, Phys. Rev. Lett. {\bf 116}, 153002 (2016).
\bibitem{Fleischhauer2005} M. Fleischhauer, A. Imamoglu, J. P. Marangos, Electromagnetically induced transparency: Optics in coherent media, Rev. Mod. Phys. {\bf 77}, 633 (2005).
\bibitem{Weis2010} S. Weis, R. Rivière, S. Deléglise, E. Gavartin, O. Arcizet, A. Schliesser, T. J. Kippenberg, Optomechanically Induced Transparency, Science {\bf 330}, 1520 (2010).
\bibitem{Suzuki2011} H. Tanji-Suzuki, W. Chen, R. Landig, J. Simon, V. Vuletić, Vacuum-Induced Transparency, Science {\bf 333}, 1266 (2011). 
\end{thebibliography}
\end{document}